\UseRawInputEncoding
\documentclass[3p,times]{elsarticle}

\usepackage{ecrc}
\usepackage{booktabs}
\usepackage{tabularx}

\volume{00}

\firstpage{1}

\journalname{Artificial Intelligence in Medicine}

\runauth{}

\jid{procs}

\jnltitlelogo{Artif. Intell. Med.}

\CopyrightLine{2023}{Published by Elsevier Ltd.}

\usepackage{amssymb}

\usepackage[figuresright]{rotating}

\begin{document}

\begin{frontmatter}



\dochead{}

\title{Multi-organ segmentation: a progressive exploration of learning paradigms under scarce annotation}{}

\author[1,2]{Shiman LI}{}
\author[1,2]{Haoran WANG}{}
\author[1,2]{Yucong MENG}{} 
\author[1,2]{\\Chenxi ZHANG\corref{cor1}}
\author[1,2]{Zhijian SONG\corref{cor1}}
\cortext[cor1]{Corresponding authors:\\zjsong@fudan.edu.cn; Chenxizhang@fudan.edu.cn}


\address[1]{Digital Medical Research Center, School of Basic Medical Science, Fudan University, Shanghai {\rm 200032}, China}
\address[2]{Shanghai Key Lab of Medical Image Computing and Computer Assisted Intervention, Shanghai {\rm 200032}, China}

\begin{abstract}
Precise delineation of multiple organs or abnormal regions in the human body from medical images plays an essential role in computer-aided diagnosis, surgical simulation, image-guided interventions, and especially in radiotherapy treatment planning. Thus, it is of great significance to explore automatic segmentation approaches, among which deep learning-based approaches have evolved rapidly and witnessed remarkable progress in multi-organ segmentation. However, obtaining an appropriately sized and fine-grained annotated dataset of multiple organs is extremely hard and expensive. Such scarce annotation limits the development of high-performance multi-organ segmentation models but promotes many annotation-efficient learning paradigms. Among these, studies on transfer learning leveraging external datasets, semi-supervised learning using unannotated datasets and partially-supervised learning integrating partially-labeled datasets have led the dominant way to break such dilemma in multi-organ segmentation. We first review the traditional fully supervised method, then present a comprehensive and systematic elaboration of the 3 abovementioned learning paradigms in the context of multi-organ segmentation from both technical and methodological perspectives, and finally summarize their challenges and future trends.

\end{abstract}


\end{frontmatter}



\section{Introduction}\label{sec1}
Medical image segmentation is an important task in medical image analysis and is widely applicable to images such as microscopy, X-ray, ultrasound, computed tomography (CT), magnetic resonance imaging (MRI), and positron emission tomography (PET). The tissue segmentation in the medical images can help clinicians evaluate the lesion size and morphology, and visualize different anatomical regions, which plays essential roles in computer-aided diagnosis, surgical simulation, treatment planning, and image-guided interventions \cite{1,2}. For instance, the success of radiotherapy treatment relies on accurate multi-organ segmentation. The accurate delineations of the organ at risk (OAR) and the target tumor contribute to precision radiotherapy treatment planning (RTP) and improvement of radiotherapy outcomes \cite{3,4}. In radiotherapy, to deliver a proper therapeutic dose to the target tumor while protecting OARs, multiple OARs and target tumor have to be delicately delineated by experienced experts, which is a highly time-consuming and labor-intensive task \cite{5,6}, which may cause delays in treatment and fall far short of clinical needs. In addition, the delineation results vary among radiologists \cite{7}, and such inter-observer variation causes contour differences accompanying a certain amount of dosimetry discrepancies \cite{nelms_variations_2012}, which may cause complications and affect the prognosis \cite{zhang_comparison_2020}.To this end, computer-assisted rapid, precise, and automated segmentation methods for multi-organ segmentation with short waiting time, high reproducibility, and consistency \cite{thor_using_2021,kawahara_stepwise_2022} have proven clinical feasibility with substantial academic and clinical value \cite{cerrolaza_computational_2019,lei_deep_2020}.

In recent years, thanks to the fast development of artificial intelligence in medical image processing, deep learning methods represented by convolutional neural networks (CNN) have dominated medical image segmentation \cite{cicek_3d_2016,bui_3d_2017,milletari_v-net_2016,zhou_unet_2020}. Likewise, in multi-organ segmentation, deep learning methods have achieved promising performance via supervised learning \cite{wang_separated_2022,taghanaki_combo_2019,cros_managing_2021,guo_organ_2020,tang_clinically_2019}. In these works, it is essential to provide pixel-level fine-grained annotations for each organ.

Despite the achievements in multi-organ segmentation of medical images using fully supervised deep learning methods, they all face the same dilemma: annotation scarcity under deep learning techniques, which often leads to model overfitting and insufficient generalization performance for unseen clinical data. 

In reality, obtaining densely annotated multi-organ data is extremely expensive and challenging \cite{hesamian_deep_2019}, for such reasons: 1) Experienced professionals are required for annotation, and organs in different part of the body calls for different experts with specific knowledge. 2) Due to the privacy concerns unique to medical images, extensive clinical data are kept in-house and away from public academic research. 3) Due to complex anatomical structures, with varying shapes, sizes, and contrasts in medical images, spatial occlusion, and blurring boundaries between adjacent organs make the accurate annotation of multi-organ an extremely hard and tedious task \cite{sadad_review_2021}. Arguably, annotation scarcity serves as a bottleneck limiting the performance of deep learning-based multi-organ segmentation models.

To alleviate the problem of annotation scarcity in multi-organ segmentation, several promising learning paradigms have been proposed and drawn attention. These new learning paradigms can be roughly divided into three categories: 1) Learning using external labeled datasets, such as transfer learning \cite{karimi2021transfer}. In transfer learning, fine-grained multi-organ annotation for a target task is not necessary and a pre-trained model on the external dataset is reused for a similar multi-organ segmentation task. 2) Learning using unlabeled datasets, such as semi-supervised learningg \cite{chapelle2009semi}. Semi-supervised learning only uses a small portion of dense annotation and a large amount of unlabeled data for model training. 3) Learning by fusing multiple partially annotated datasets, such as partially-supervised learning \cite{zhou_prior-aware_2019}. Partial supervision can use multiple single-organ or partial-organ datasets as valid annotations to segment multiple organs of inputs simultaneously. The above learning paradigms are ushering in new research directions for deep learning-based segmentation of multi-organ images.

Existing reviews on multi-organ segmentation have mainly focused on the area of fully supervised learning \cite{lei_deep_2020,ilesanmi_organ_2022,fu_review_2021,cerrolaza_computational_2019}, while reviews on annotation scarcity are still absent. Recently, the exploration of annotation-efficient methods has received increasing attention in medical image segmentation. Cheplygina et al. \cite{cheplygina2019not}, Zhang et al. \cite{zhang2020survey} and Tajbakhsh et al. \cite{tajbakhsh_embracing_2020} investigated non-fully supervised medical image segmentation methods including weakly supervised, transfer learning, active learning, and semi-supervised learning paradigms, etc. Ma et al. \cite{ma_abdomenct-1k_2021} established new segmentation benchmarks for semi-supervised, weakly supervised, and continuous learning to advance the unsolved problem that the existing satisfactory performance may not generalize on more diverse datasets. It can be observed that as new technologies continue to emerge, new learning paradigms gain wider popularity. Hence, a systematic review of studies on annotation-efficient multi-organ segmentation is urgent and valuable.

In this review, we comprehensively summarize different frameworks of the typical fully supervised learning and challenges in fine-grained segmentation for multi-organ segmentation tasks, and survey relevant works on transfer learning, semi-supervised learning, and partially supervised learning from the perspective of annotation scarcity problem. To our best knowledge, we are the first to conduct a review of partially supervised learning paradigms. We search by google scholar with the keywords such as semi-supervised learning, transfer learning, domain adaptation, partially supervised, and multi-organ segmentation, and got over 200 initial articles. After filtered by investigating abstract, we obtained 126 highly relevant works and our review contained 201 references.

\begin{figure}
\centering
\includegraphics[width=1.0\textwidth]{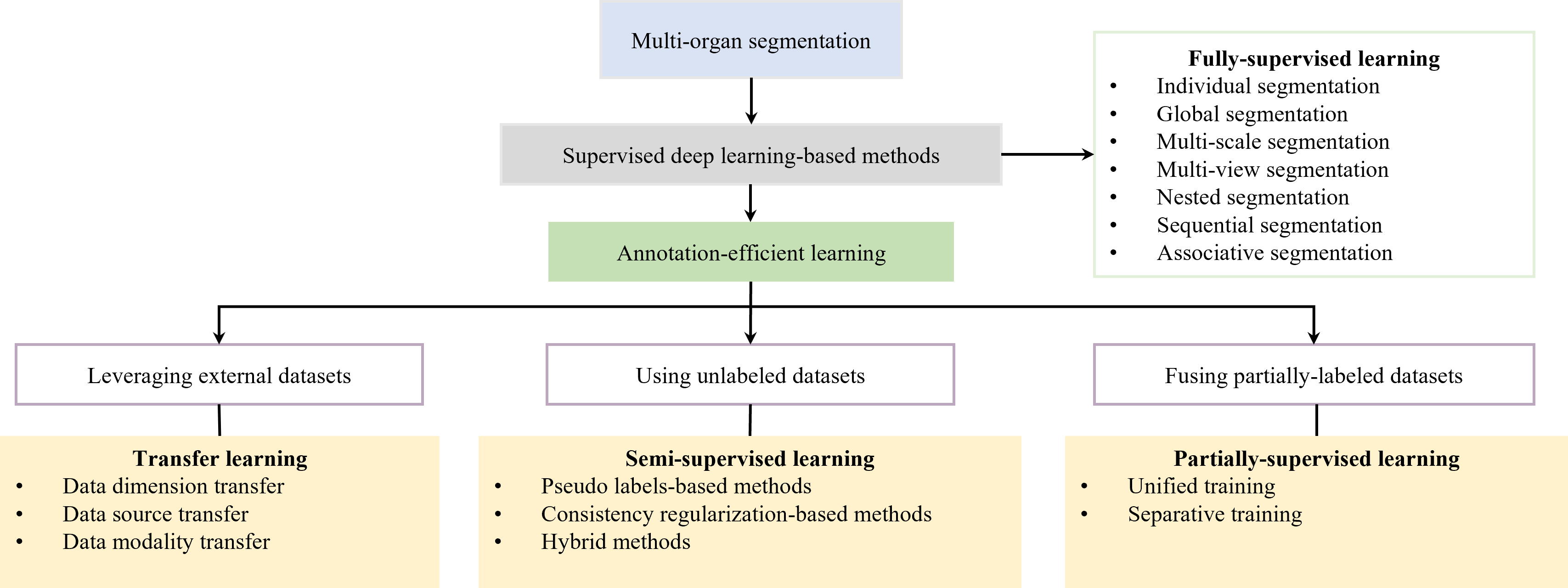}
\caption{Organization of this review paper. We roundly present the basic segmentation mode of multi-organ segmentation in the fully supervised learning methods and categorize the annotation-efficient learning methods into three classes to embrace the scarce annotations. Each annotation-efficient learning will be categorized and described based on its specific technical details.}\label{<figure1>}
\end{figure}

It should be noted that, on the definition of multi-organ, except for the OARs of radiotherapy, we also encompass organs \& tumors or multiple tissues, in a word, we adopted segment multiple anatomical structures simultaneously as the task of multi-organ segmentation in this paper. For the sake of grouping the overarching ideas, we skip the illustration of particular challenges for different regions, such as head and neck \cite{daisne_atlas-based_2013,lim_use_2016} or abdomen \cite{li_overview_2021,qayyum_automatic_2020}, but explore the general problem of multi-organ segmentation under deep learning from the perspective of solving annotation scarcity.
The rest of the paper is structured as follows. Section \ref{sec2} introduces the general modes and challenges of multi-organ segmentation, and Section \ref{sec3} describes a detailed review of transfer learning (Section \ref{subsec1}), semi-supervised learning (Section \ref{subsec2}), and partially supervised learning (Section \ref{subsec3}). In Section \ref{sec4}, we discuss future trends of the three learning paradigms, and we conclude the whole paper in Section \ref{sec5}. And the organization of the whole review is shown in Figure \ref{<figure1>}.

\section{Overview of multi-organ segmentation \& problem analysis}\label{sec2}
The segmentation of multiple organs is of great clinical significance, providing quantitative information for the diagnosis and classification of diseases \cite{telford_mr_2014,CONZE2021102109,KHAN2022102231}, and is an indispensable part of radiotherapy. Currently, deep learning-based image segmentation methods are commonly classified into convolutional neural networks (CNN) \cite{zhou_unet_2020}, fully convolutional networks (FCN) \cite{long2015fully}, and generative networks (GAN) \cite{mondal_few-shot_2018} according to the structure of the network \cite{hesamian_deep_2019}. And recently, Transformer \cite{vaswani2017attention}, which has been widely used in natural images, has also been introduced into medical organ segmentation \cite{cao2021swin,azad_transnorm_2022}. All these methods have achieved good results in the field of multi-organ segmentation. Among them, U-Net \cite{ronneberger_u-net_2015} and its variant structures \cite{cicek_3d_2016,bui_3d_2017,milletari_v-net_2016,zhou_unet_2020,isensee2021nnu}, the classic fully convolutional networks, have become the prevailing method for the multi-organ segmentation of medical images owing to their powerful capabilities. Alternatively, if classified from the training procedure, apart from the basic end-to-end segmentation process, the two-stage approaches based on localization-segmentation \cite{wang_organ_2019,dogan_two-phase_2021,gu_fusing_2021,zhao_knowledge-aided_2019} and the cascade optimization \cite{roth_application_2018,cao_cascaded_2021,cao_2d3d_2022,zhong_boostingbased_2019} approaches are also common in multi-organ segmentation for the enhancement of delineation accuracy.

Some approaches simplify multi-organ segmentation to multiple independent single-organ segmentation tasks \cite{tang_clinically_2019,wang_organ_2019,yu2018recurrent,larsson_robust_2018}, which has been experimentally verified to be effective for localizing and segmenting the multi-organ \cite{senkyire2021supervised} and some challenging organs, such as the pancreas \cite{ma_abdomenct-1k_2021}. Yet, such approaches ignore the inter-organ spatial correlation, therefore, how to make full use of such contextual information, i.e., multi-organ anatomical prior information, is an inescapable problem in multi-organ segmentation \cite{liu_context-aware_2022}. Following the definition of the categories of segmentation methods in the review by Cerrolaza et al. \cite{cerrolaza_computational_2019}, we classify the common deep learning-based multi-organ segmentation methods using context information into six different categories similarly. And it is worth mentioning that the specific scope of our survey differs from the review of Cerrolaza et al. \cite{cerrolaza_computational_2019} that focuses on traditional machine learning methods.

\textbf{Global segmentation} can be achieved by directly inputting medical images that contain multiple targets into the model for simultaneous segmentation \cite{park_development_2020,bobo_fully_2018,chen_fully_2020}. The whole-volume input preserves complete intra- and inter-organ anatomical information,  which can assist in localization and segmentation \cite{zheng_anatomically_2021}. However, for the full-volume input, it is difficult to extract the full inter-organ contextual information due to the limited receptive field of the convolutional kernel. Some sub-modules are proposed to exploit the structural contextual information, such as dilated convolution, pyramidal architecture, etc. \cite{heinrich_obelisk-net_2019,bahdanau2014neural,chen_novel_2021,zhu_anatomynet_2019} and other ideas are to make use of spatial attention mechanisms \cite{qayyum_automatic_2020,cao_cascaded_2021,xu_3d_2021,gou_self-channel-and-spatial-attention_2020,tang_spatial_2021}. The advantages of global segmentation are the simplicity of the training process and the fact that the global model can impose strong anatomical constraints for the segmentation of organs suffering from data corruption (missing, artifacts, etc.). Nevertheless, limited by the computational capability, researchers usually crop the image in preprocessing stage based on statistical and anatomical priors \cite{zhang2021efficient,roth_hierarchical_2017,weston_automated_2019,zhu_3d_2018} or coarse segmentation localization so as to further improve the foreground proportion. It is crucial for global segmentation to balance the computational consumption and the resolution-constraint segmentation accuracy.

\textbf{Multi-scale segmentation} improves segmentation performance by incorporating information or predictions under different resolutions \cite{gibson_automatic_2018,kamnitsas_efficient_2017,roth_multi-scale_2018,xue_segan_2018}. Multi-scale exploration is similar to the perception of human vision, utilizing diverse detailed information of different resolutions. At low resolution, we can easily acquire the overall inter-organ relationships, while the specific details of each organ are better examined at high resolution. Existing multi-scale studies have been conducted by fusing multi-resolution inputs \cite{wang2021patch} or multi-level feature information \cite{zhao_mss_2020} to better segment organs of highly varying sizes simultaneously. Moreover, a deeply-supervised approach \cite{lee_deeply-supervised_2015} can also optimize the segmentation by the combination of multi-scale outputs \cite{li_3d_2020}.

\textbf{Multi-view segmentation} enables more accurate spatial localization by using multi-view data to impose anatomical information constraints, which can effectively alleviate the problems of occlusion and confusion \cite{li_multi-dimensional_2022,wang_abdominal_2019,setio_pulmonary_2016}. Multi-view segmentation can model spatial information with less computational consumption by constructing multi-view attention modules with inputs from transverse, coronal, and sagittal planes or more \cite{yang_anatomy-guided_2022,perslev_one_2019}. Moreover, different views enable spatial-anisotropy modeling to mitigate the effects of various resolutions of different directions and collaboratively utilize high-resolution information within slices and inter-slice contextual information \cite{wang_abdominal_2019}. However, this method draws the disadvantage of training multiple separate models of different views.

\textbf{Nested segmentation} is a hierarchical segmentation approach with models training for both global and local organs respectively \cite{ren_interleaved_2018,zhang_slice_2021,guo_organ_2020,zhong_boosting-based_2019}. Global organ segmentation network can impose topological constraints on anatomical context information for multi-organ segmentation, and the networks for sub-organs can further refine segmentation results. The nested segmentation allows coarse-to-fine segmentation and interaction of global and local information, where the construction of hierarchical objects can be built on spatial or physiological connections. Besides the spatial relationship, joint physiological associations can be helpful to enhance the segmentation accuracy of lesions and affected organs \cite{chen_fully_2020}. However, in nested segmentation, the training complexity and time increase significantly with the number of hierarchical layers.

\textbf{Sequential segmentation} is a way of segmenting organs from easy to hard, which can help boost the prediction of hard-to-segment organs and maximize overall segmentation accuracy\cite{zhao_coarse--fine_2022}. In clinical practice, radiologists often use easily distinguishable structures (e.g., bones) as a reference to assist in outlining difficult targets (e.g., soft tissues of the brain). Similarly, the order of organ processing in sequential segmentation is also often determined based on difficulty, and such processing benefits from the fixed spatial prior \cite{zhao_coarse--fine_2022,guo_organ_2020}. For example, combining large organ prediction in the sequential segmentation can effectively guide the small adjacent organ delineation, such as feeding with the segmentation of kidneys can help contour the adrenal gland with the help of cascade training \cite{zhao_coarse--fine_2022}. In this case, sequential segmentation concludes locations for the subsequential task which also does good for organ-specific tumor identification, but the effect heavily depends on the accuracy of the preceding sequence with long training and inference time.

\textbf{Associative segmentation} delineates different organs individually but not independently by constructing bridges between models to provide anatomical interaction for simultaneous segmentation of multi-organ \cite{you_3d_2021,cros_managing_2021}. For example, plenty of symmetric structures in the human body (kidney, optic nerve, etc. \cite{ren_interleaved_2018,harms_automatic_2021}) can be optimally segmented by shape similarity and spatial symmetry constraint, and the lesion region can also be identified by disparity. Furthermore, an association of asymmetric organs is equally effective for better segmentation performance, which was experimental confirmed in the segmentation of highly variable and non-directly connected organs such as the pancreas and spleen \cite{fu_domain_2020}. Associative segmentation can add flexible inter-organ constraints to the models of each organ, allowing for higher segmentation accuracy. However, multiple coupled segmentation networks pose challenges to models’ convergence and training.

Although the above common multi-organ segmentation methods effectively leverage contextual information, there are still two issues that deserve improvement. (1) Low segmentation accuracy of tiny or low-contrast organs. Segmentation of such organs is very challenging, caused by the large variation in organ size and intensity distribution and class imbalance~\cite{cros_managing_2021,li_analyzing_2021}. Apart from the two-stage approach mentioned above~\cite{gao_focusnet_2019,gao_focusnetv2_2021,johnson_survey_2019}, the additional loss function is also a tool to reduce the class imbalance, but the gain is low when dealing with too many organs~\cite{taghanaki_combo_2019,shi_marginal_2021,sudre_generalised_2017}. Alternatively, organ-specific preprocessing is beneficial for problem solving~\cite{wang_learning-based_2019,zhao_knowledge-aided_2019}, such as intensity regularization~\cite{zhou_normalization_2019} and bias correction~\cite{lei_learning-based_2019}, but this requires additional prior information and is not generalizable across organs. (2) Large computational resource occupation. Some existing methods sacrifice training efficiency and network simplicity for higher accuracy, making the hardware requirements (e.g., memory, computational speed) unsuitable for realistic clinical practice~\cite{belal_deep_2019,ali_state---art_2021,qiu_automatic_2021}.

In general, extended from earlier single-organ segmentation methods, multi-organ segmentation methods have gradually combined the robustness of global multi-organ modeling with the flexibility and versatility of individual segmentation strategies to balance accuracy and computational efficiency. By incorporating context information, models are able to cope with the effects of overlap, gaps, or topological deformation, thus achieving a more precise representation of complex human anatomy.  In recent years, deeper networks have been proposed and have shown stronger feature extraction capabilities. However, due to the limited quality, size, and consistency of the publicly available datasets, network overfitting, and over-parameterization have intensified resulting in insufficient generalization more so than before~\cite{chen_deep_2020,bagherzadeh2019review,lee_semi-supervised_2020}. This makes annotation-efficient algorithms a key to breaking the data scarcity bottleneck and thus enabling performance improvement in the future.

\section{Learning paradigms under annotation scarcity}\label{sec3}

\subsection{Leveraging external annotated datasets - Transfer learning}\label{subsec1}

Rarely do we access enough data and labels for the target domain with respect to multi-organ segmentation, in which case, we can segment target datasets via transfer learning. Typical transfer learning involves pre-training a model with a large external annotated dataset and then fine-tuning the model on the target dataset. The basic concept of this method comes from the fact that the same organs from various datasets may have similar anatomical morphology and surrounding structures in medical image segmentation. Such learnable knowledge can be transferred to the segmentation of the target domain data, or serve as auxiliary information contributing to performance enhancement. Inevitably, medical images from different datasets involve distribution gaps across various modalities, scanners, protocols, and clinical sites, etc., degrading the performance. How to mitigate the impact of such domain shift is an inherent problem of transfer learning. This section elaborates on the existing studies in terms of different domain adaption modes, including dimensions, sources, and modalities. We review and comparatively present the representative work on multi-organ segmentation using transfer learning in Table \ref{tab1}.

\begin{table*}[t]
  \footnotesize
\begin{center}
\caption{Overview of multi-organ segmentation work leveraging external datasets.}\label{tab1}%
\footnotesize
\begin{tabular}{@{}m{1cm}<{\centering}m{1.5cm}<{\centering}m{1.5cm}<{\centering}m{2.2cm}<{\centering}m{8cm}<{\centering}@{}}
\hline
Reference & Segmentation mode & Site    & Domain adaption mode                               & Transfer technique                                                                                               \\
\hline
\cite{yu2018recurrent}  & Individual       & Abdomen               & 2D $\rightarrow$ 3D                                          & Axis transform; pseudo-label   generation                                                                                            \\
\cite{liu20183d} & Multi-view        & Liver and tumor       & 2D $\rightarrow$ 3D                                          & Anisotropic hybrid network                                                                                                           \\
\cite{zheng2020annotation} & Multi-view        & Heart                 & 2D $\rightarrow$ 3D                                          & Pseudo-label generation                                                                                                              \\
\cite{dong2018unsupervised} & Global            & Chest                 & Public $\rightarrow$ clinical                                & Adversarial training of shared   segmentor with discriminator on predicted masks                                                     \\
\cite{ma2019neural} & Multi-view        & Heart                 & Multi-devices                                                & Slices generation by neural   style transfer                                                                                         \\
\cite{huang_cross-dataset_2021} & Global            & Whole body organs     & Public $\rightarrow$ clinical                                & Filtered back projection-driven   domain adaptation                                                                                  \\
\cite{fu_domain_2020}  & Global            & Abdomen               & Clinical $\rightarrow$ public                                & Shared knowledge transfer   of fixed spatial location by solving puzzle task; resolution normalization by   super-resolution network \\
\cite{ma_rapid_2022} & Global            & Abdomen               & Public $\rightarrow$ clinical                                & Human-in-the-loop                                                                                                                    \\
\cite{zhang_task_2018} & Global            & Chest                 & CT $\rightarrow$ X-ray                                       & Task-driven generative   network                                                                                                     \\
\cite{jiang_unpaired_2022} & Global            & Abdomen               & CT $\rightarrow$ MR                                          & Translation by CycleGAN                                                                                                              \\
\cite{chen2020unsupervised} & Global            & Heart                 & BSSFP $\rightarrow$ LGE                                      & Translation by CycleGAN                                                                                                              \\
\cite{hong_source-free_2022}  & Global            & Abdomen               & CT $\rightarrow$ MR                                          & Style-compensation generation;   statics-guided feature adaption                                                                     \\
\cite{wang_towards_2022} & Global            & Heart, Abdomen        & CT $\rightarrow$ MR                                          & Appearance transformation   and representation transfer by GAN                                                                       \\
\cite{jiang_unified_2020} & Global            & Abdomen               & CT $\leftrightarrow$ T1w $\leftrightarrow$ T2w & Generation by VAE   separating style and content                                                                                     \\
\cite{xie_unsupervised_2022} & Global            & Heart, Abdomen, Brain & CT $\leftrightarrow$ MR                               & Learning domain-invariant   and domain-specific components                                                                           \\
\cite{valindria_multi-modal_2018} & Global            & Abdomen               & CT $\leftrightarrow$ MR                               & Learning shared knowledge   by vary structures in Figure \ref{<figure2>}                                                                           \\
\cite{dou_pnp-adanet_2018} & Global            & Heart                 & CT $\leftrightarrow$ MR                               & A shared decoder to produce   consistent high-level semantic embeddings                                                              \\
\cite{yang_toward_2022} & Global            & Heart, Abdomen        & CT $\leftrightarrow$ MR                               & A single shared network with attention module to learn structured semantic consistency  \\
\bottomrule
\end{tabular}
\end{center}
\end{table*}

\subsubsection{Dimension transfer}\label{subsubsec1}

Since the annotation of 2D data is easier than that of 3D data, using 2D annotated data for the segmentation of 3D data is an effective way to reduce the annotation workload. Yu et al.~\cite{yu2018recurrent} proposed a recurrent saliency transformation network, which transfers the third dimension of the image into a temporal axis to generate 3D pseudo-labels from 2D prediction, and converts the predicted probabilities into spatial weights with the aim of improving the results of medium and small organs via multi-stage cascade segmentation. However, this method can hardly capture 3D context information. Meanwhile, in dimension transfer, the intra-slice resolution is much higher than the inter-slice resolution, in which the direct employment of 3D convolutions may lead to distorted extraction of image features. To tackle such issues, Liu et al.~\cite{liu20183d} proposed an anisotropic hybrid network to migrate convolutional features learned from 2D images to anisotropic 3D images segmentation. The network achieved advanced results with its powerful extraction of intra-slice information and effective modeling of inter-slice information. Additionally, Zheng et al.~\cite{zheng2020annotation} proposed to select the most representative slices for annotation and performed a 3D segmentation with these few annotation slices. This selection strategy used 20\% of labeled slices to obtain results comparable to those obtained with 100\% of labeled slices. In this work, they used a multi-view model to integrate contextual information and the pseudo label to implement dimension transfer.

\subsubsection{Source transfer}\label{subsubsec2}

Data from different sources often differ in data distribution due to imaging parameters, imaging devices, etc. With a view to applying the existing research achievements based on public datasets to clinical datasets, we could pre-train the model on public datasets, then transfer it to the segmentation task on new clinical private data. Dong et al.~\cite{dong2018unsupervised} argue that the prediction masks have domain-invariant anatomical similarity, and therefore implemented adversarial training on the prediction masks by discriminator to segment the chest organs for cardio-thoracic ratio estimation. Apart from the adversarial-based approach, Ma et al.~\cite{ma2019neural} proposed a neural style transfer approach that reduces the data inconsistency between sources to minimize the variation in image attributes (including brightness, contrast, texture, etc.) and integrates global information using atrous convolution and pyramidal pooling. Huang et al.~\cite{huang_cross-dataset_2021} enforced FBP (Filtered Back Projection)-driven domain adaptation techniques to minimize the gap across different CT image datasets, and fused the contextual information through global organs’ registration. Fu et al.~\cite{fu_domain_2020} employed the relatively fixed spatial location between organs as potentially transferable shared knowledge for domain adaptation of multi-center data. It learns spatial relationships by solving a puzzle task and normalizes the spatial resolution of different sources through a super-resolution network, yielding an average of 19.6\% improvement in dice similarity coefficient (DSC) on five publicly target dataset via transfer learning from a clinical labeled dataset.

Besides, human-in-the-loop is an important idea to facilitate the clinical usage of computer-aided technology, where rapid transfer of AI models from one dataset to another can be supplemented with human intervention. Ma et al.~\cite{ma_rapid_2022} proposed a novel and general human-in-the-loop scheme to transfer a model trained with small-scale labeled data to a multi-organ segmentation task of large-scale unannotated CT images, which not only improves the segmentation accuracy but also reduces the manual annotation cost from 13.87 minutes to 1.51 minutes per sample, demonstrating its potential for clinical application.

\begin{figure}
\centering
\includegraphics[width=0.8\textwidth]{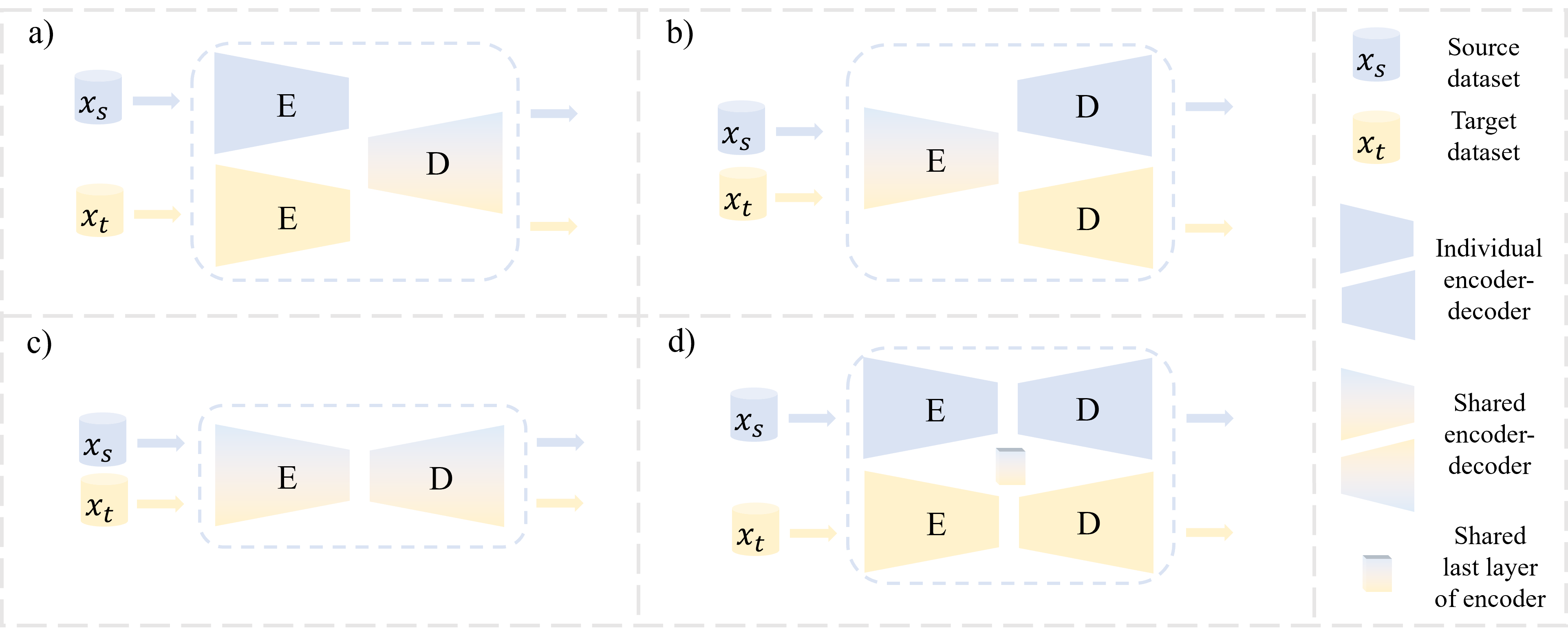}
\caption{The diagram of dual-stream models in the work by Valindria et al.~\cite{valindria_multi-modal_2018}. a) modality-specific encoder and shared decoder. b) shared encoder and modality-specific decoder. c) shared encoder and decoder. d) different streams in the encoder, shared last layers of the encoder, and modality-specific decoder.}\label{<figure2>}
\end{figure}

\subsubsection{Modality transfer}\label{subsubsec3}

Medical images from diverse modalities have different characteristics and complementary information. The multimodal medical images and their cross-modality information can not only support radiologists in making more accurate diagnoses but have also proven effective in automatic AI segmentation~\cite{sohail_unpaired_2019,zhang_multi-contrast_2022,xin_multi-modality_2020}. Cross-modality knowledge transfer is the major emphasis of transfer learning for medical images, and it's essential to implement domain adaption by learning latent shared representations or translating images from one modality to another. The presence or absence of target domain annotation often calls for different research ideas.

\textbf{Without target domain annotations: }When the target domain is unlabeled, unsupervised learning is performed on the target modality. In this case, segmentation of the target domain often relies on segmentation networks trained using source domain annotations~\cite{bousmalis_unsupervised_2017}. Generative adversarial network (GAN)~\cite{goodfellow2020generative} and cycle-consistent generative adversarial network (CycleGAN)~\cite{zhu_unpaired_2020}, trained in an adversarial mode, are the commonly used methods to address domain shifts across modalities with generators and discriminators. By encouraging the generators to produce synthetic images that are sufficient to confuse the discriminators, the generators can translate images from one domain to another. Zhang et al.~\cite{zhang_task_2018} used CT images to synthesize X-Ray like digitally reconstructed radiographs through a task-driven generative network, and then trained an X-Ray depth decomposition model based on these images~\cite{albarqouni_x-ray_2017}, which was transferred to real X-ray images for multi-organ segmentation. Jiang et al.~\cite{jiang_unpaired_2022} proposed cross-modality educed distillation (CMEDL) based on a teacher-student network, where CycleGAN was used to achieve MRI-to-CT translation, and conducted knowledge distillation on CT and translated MRI images. Similarly, Chen et al.~\cite{chen2020unsupervised} used CycleGAN for translation between MR and CT and used an encoder shared with the generator to prevent encoder overfitting in CT heart segmentation. 

However, it is difficult to achieve a satisfactory translation with a simple whole-image reconstruction, due to the loss of valuable detail information of target modality. Therefore, certain adaptation techniques can improve the translation quality thus support better segmentation performance of generation-based transfer learning. Hong et al.~\cite{hong_source-free_2022} proposed feature map statistics-guided model adaptation combined with entropy minimization to accomplish knowledge transfer and reliable segmentation, and implemented style compensation using CycleGAN to significantly improve the segmentation accuracy of the left kidney and spleen, and the effectiveness was validated experimentally for domain adaptation from both CT to MR and its inverse. Wang et al.~\cite{wang_towards_2022} explored collaborative appearance and semantic adaption with a characterization transfer module to alleviate the appearance divergence of medical lesions across domains and a representation transfer module to diminish the domain-wise distribution gap of underlying semantic knowledge. Alternatively, the disentanglement approach can extract domain-invariant content and domain-specific styles making the generation more robust and reducing catastrophic forgetting and model collapse~\cite{liu_detach_2018,huang_multimodal_2018,lee_diverse_2018}. Jiang et al.~\cite{jiang_unified_2020} separated image style and content by variation auto-encoder~\cite{kingma2013auto} (VAE), which converts image style to latent style code and generates multi-modal images for the content of the input image. Xie et al.~\cite{xie_unsupervised_2022} learnt domain-invariant and domain-specific anatomical components of images by disentanglement learning, and enhance the target domain segmentation performance by adversarial learning and pseudo labeling in self-training to implicitly achieve feature alignment in anatomical space.

\textbf{With target domain annotations:} When the target domain is annotated, data from other modalities often serve as supplementary information to improve the segmentation results~\cite{valindria_multi-modal_2018,xin_multi-modality_2020}. Unlike unsupervised domain adaptation, domain adaptation techniques are more focused on learning shared knowledge for meaningful information transfer when both datasets are labeled. Valindria et al.~\cite{valindria_multi-modal_2018} explored the performance impact of 4 different encoder-decoder settings on extracting shared latent representations in an abdominal multi-organ segmentation scenario, as shown in Figure. 2. The experimental results show that dual streams of Fig.2 (d) outperform the other structures, besides, cross-modal information can improve the segmentation performance of mutable structures such as the spleen. Dou et al.~\cite{dou_pnp-adanet_2018} trained a heart segmentation network consisting of two domain-specific encoders and a shared decoder, enforcing the decoder to produce similar high-level semantic embeddings for both domain images by adversarial training. Yang et al.~\cite{yang_toward_2022}, on the other hand, explored a single shared network to integrate multi-modality information, experimenting with a single transformer and a well-designed external attention module to learn structured semantic consistency (i.e., semantic category representations and their correlations). And the method achieved competitive performance even with extremely limited labeled samples.

\subsection{Using unannotated datasets – Semi-supervised learning}\label{subsec2}

In medical images, unlabeled data are often more accessible than labeled data. When labeled data are scarce, the inclusion of unlabeled data has been shown to help improve model generalization and the segmentation performance of medical images~\cite{zhang2020survey,jiao_learning_2022,MEYER2021102073}. Semi-supervised learning is trained using only a small set of labeled data and large-scale unlabeled data, where the key to the investigation is to make sufficient utilization of unlabeled data~\cite{chapelle2009semi}. In this section, observing the existing mainstream studies, we categorize the approach and show the typical workflows of semi-supervised learning in Fig.3, including the pseudo label-based approach, consistency regularization approach, and hybrid approach.

\begin{figure}
\centering
\includegraphics[width=1.0\textwidth]{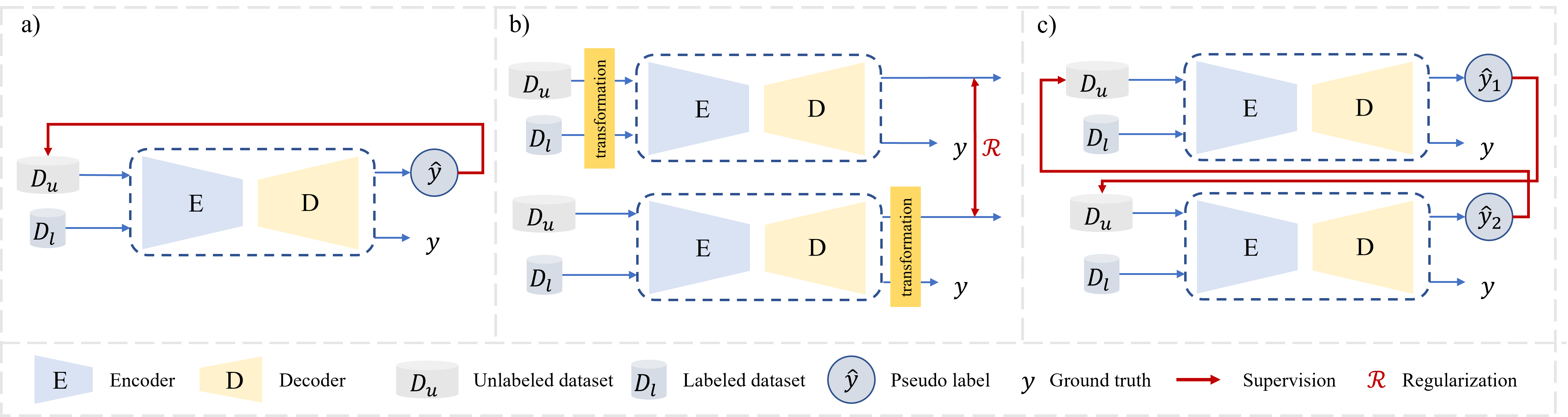}
\caption{Typical workflows semi-supervised methods. a) a basic workflow of the pseudo-labeling method, which uses the prediction of unlabeled data for pseudo label-based supervision; b) an example of consistency-based training using transformation consistency, in which the predictions from two networks are supervised by a consistent regularization; c) an example of a hybrid method using cross pseudo supervised, which construct a regularization by exchanging pseudo labels. }\label{<figure3>}
\end{figure}

\subsubsection{Pseudo label-based methods}\label{subsubsec1}
The semi-supervised method using pseudo labels was first proposed by Lee et al.~\cite{lee2013pseudo} and has been developed as a classical algorithm for semi-supervised for its simple idea and effective performance. As shown in Figure \ref{<figure3>} (a) the model trained by labeled data can be used to predict the pseudo label of unlabeled data, which is fed together with labeled data to co-train the model in an iterative manner, and make the model achieves promising results on validation set outperforming the training using only labeled data. The such a pseudo-labeling approach has a high universality, but its effectiveness is restricted by the quality of the pseudo labels generated from the unlabeled data. And the related works mentioned below differ in the way of pseudo label generation and the treatment of noisy labels and are listed in Table \ref{tab2}.

\begin{table}[h]
\begin{center}
\caption{Summary of semi-supervised medical segmentation methods using pseudo labels.}\label{tab2}%
\footnotesize
\begin{tabular}{@{}m{1cm}<{\centering}m{1.5cm}<{\centering}m{1cm}<{\centering}m{5cm}<{\centering}m{5.5cm}<{\centering}@{}}

\toprule
Reference & Segmentation mode & Site    & The generation of pseudo labels                               & Handling of noisy pseudo label                                                                                               \\
\midrule
\cite{zhou_semi-supervised_2019} & Multi-view        & Abdomen & Ensemble segmentation model of three views \& Majority voting & /                                                                                                                            \\
\cite{zhao_data_2019}  & Global            & Brain   & Average of multi-transformed segmentation results             & /                                                                                                                            \\
\cite{min_two-stream_2019} & Global            & Heart   & Two-stream mutual attention network                           & Filtered by the difficulty level obtained from hierarchical distillation, including data distillation and model distillation \\
\cite{nie_asdnet_2018} & Global            & Pelvis  & Single adversarial segmentation model                         & Masked by the confident map predicted with discriminator                                                                     \\
\cite{xia_uncertainty-aware_2020} & Multi-view        & Abdomen & Multi-view co-training network                                & Uncertainty-weighted label fusion                                                                                            \\
\cite{ma_abdomenct-1k_2021}  & Global            & Abdomen & Noisy student network                                         & Update pseudo labels with training iterations    \\
\bottomrule
\end{tabular}
\end{center}
\end{table}
Zhou et al.~\cite{zhou_semi-supervised_2019} proposed an iterative model for the segmentation of 16 structures in contrast-enhanced abdominal CT, and the pseudo labels are generated by maximum voting from models of axial, coronal, and sagittal views, which eventually brings a 4\% improvement in DSC due to the usage of unlabeled data. In the work by Zhao et al.~\cite{zhao_data_2019}, the pseudo label is derived by data distillation, which averages the prediction masks from random augmentations of the same image to produce the final pseudo label, and the proposed method also yield a 1.6\% DSC improvement compared to training without unlabeled image.

Nevertheless, the above methods have obvious limitations, mainly in treating the pseudo-labeled and expert-labeled annotation equally, without considering that false segmentation in pseudo label may bring noise into the model training and thus degrade the model performance~\cite{lee_semi-supervised_2020,rizve2021defense,yao_enhancing_2022}. To explore such an issue, an important approach is to estimate the reliability of pseudo labels. Min et al.~\cite{min_two-stream_2019} proposed a dual-stream network and the two independent networks can discriminate the difficulty of each pixel by the consistency of prediction for the same input of K transformations and two models, in which case the hard and easy pixels with inconsistent classes will be shielded from the overall loss calculation. The network generates pseudo labels by hierarchical distillation including both model distillation and data distillation and demonstrates the noise-resilient of the suggested method on a full-heart segmentation dataset under varying noise levels of labeled images. Instead, Nie et al.~\cite{nie_asdnet_2018} proposed a region-attention-based strategy based on the confidence map gained from an adversarial confident network, then used confident region obtained by setting threshold as a mask to select reliable prediction of unlabeled data for pseudo label-based supervision, this method demonstrated their superiority over fully supervised methods on multiple datasets. On the other hand, Xia et al.~\cite{xia_uncertainty-aware_2020} proposed a co-training framework to predict segmentation masks by multi-view parallel network predictions. During training, multiple inference results with random dropout~\cite{gal_dropout_2016} can be used to estimate uncertainty served as sample weights, alleviating the effect of noisy pixel labels and achieving significant performance improvements on multi-organ datasets.

Beyond reliability estimation, many works have likewise explored the utilization of noisy pseudo labels. Ma et al.~\cite{ma_abdomenct-1k_2021} established a new semi-supervised segmentation benchmark for the abdominal multi-organ dataset using the Noisy Student framework~\cite{xie2020self}. The noisy student can self-correct the pseudo labels in the training phase, and the student model includes more data augmentation and noise and its prediction will play the role of the new pseudo labels to iterate and update the student model itself. Different from basic pseudo labels training, not only the model but also the pseudo labels are continuously optimizing in the multiple iterations. Such a simple idea demonstrated its powerful generality and achieved efficient performance, with an experimental illustration that errors in difficult and diseased organ segmentation can be gradually corrected using an increasing amount of unlabeled data.

\subsubsection{Consistency regularization-based methods}\label{subsubsec2}

Without pseudo-labeling, the unlabeled data in semi-supervised learning rely on setting up an unsupervised loss together with the supervised loss for labeled data to guide the backpropagation. The most prevalent way of unlabeled loss construction is often based on the idea of consistent regularization, i.e., the model is robust to small perturbations and the outputs may stay consistent~\cite{sajjadi2016regularization,rasmus2015semi}. Inspired by the classical semi-supervised segmentation frameworks Temporal Ensembling~\cite{laine2016temporal}, Mean Teacher~\cite{tarvainen2017mean}, and UDA~\cite{xie2020unsupervised}, many semi-supervised methods have also been extended to multi-organ segmentation. Table \ref{tab3} lists some representative consistency regularization-based semi-supervised multi-organ segmentation methods which differ in perturbations and consistency construction.

\begin{table}[h]
\begin{center}
\caption{Summary of consistency regularization-based methods in semi-supervised medical multi-organ segmentation}\label{tab3}%
\footnotesize
\begin{tabular}{@{}m{1cm}<{\centering}m{1.5cm}<{\centering}m{1.5cm}<{\centering}m{5cm}<{\centering}m{5cm}<{\centering}@{}}

\toprule
Reference & Segmentation mode & Site        & Perturbations                                                   & Consistency type                                           \\
\midrule
\cite{bortsova2019semi} & Global            & Chest       & Elastic deformations                                            & Elastic transformation consistency                         \\
\cite{basak_embarrassingly_2022} & Global            & Heart       &  Mixup transformations                                           & Interpolation consistency                                  \\
\cite{luo_efficient_2021} & Multi-scale       & Nasopharynx & Pyramid network architecture                                    & Multi-scale consistency                                    \\
\cite{meng_shape-aware_2022} & Global            & Optic       & Tasks of segmentation and regression                            & Task consistency of regional and marginal consistency      \\
\cite{peng_boosting_2021} & Global            & Heart       & Typical input-level transformation: random crop, flip, rotation & Feature consistency on global and local mutual information \\
\cite{hu_semi-supervised_2022} & Global            & Nasopharynx & Teacher-student network with different inputs                   & Attention map consistency       \\
\bottomrule
\end{tabular}
\end{center}
\end{table}

Transform consistency is a common consistency regularization~\cite{li_semi-supervised_2018,li_transformation_2020}, and mostly performs in a regular workflow like that shown in Figure \ref{<figure3>} in b. This inspire Bortsova et al.~\cite{bortsova2019semi}, to use elastic transform consistency for segmentation, which constructs consistency of the final outputs from two branches to help learn transformation invariant representations and further boost the utilization of unlabeled data. And this method effectively improves segmentation results under a small number of supervised data, but yield little improvement compared to fully-supervised methods when the number is large. Similarly, Basak et al.~\cite{basak_embarrassingly_2022} conducted effective semi-supervised segmentation by interpolation transform consistency using Mixup~\cite{zhang_mixup_2018} for data transformation, which extends the training set by mixing diverse images.

More various consistency constructions have been proposed too, such Luo et al.~\cite{luo_efficient_2021} suggested uncertainty rectified pyramid consistency, a kind of multi-scale consistency, which is formulated by the MSE loss from the pyramid outputs to their average. The pyramid prediction network integrates contextual information at different scales and the predictions are rectified by the uncertainty map produced by the predictions' entropy. Also, Meng et al.~\cite{meng_shape-aware_2022} proposed a bi-consistency regularization with regional and marginal consistency to build a shape-aware generalized multi-task framework improving segmentation accuracy for both semi-supervised and weakly-supervised~\cite{zhou_brief_2018} segmentation. 

Moreover, rather than ensuring consistency at the model, task, or output level as mentioned above, many studies are now gradually focusing on feature information utilization and adding regularization constraints at the feature level. As Peng et al.~\cite{peng_boosting_2021} constrained consistency on feature-level mutual information to learn a geometrically transform invariant image representation with the help of global mutual information, and then promote spatial consistency of the feature map by local mutual information to provide smoother segmentation results. Hu et al.~\cite{hu_semi-supervised_2022} then introduced attention map consistency in an uncertainty-aware teacher-student model between the feature of the whole image or the ROI-cropped image inputs, which helps the model better locate the target and reduce the false positives of segmentation.

\subsubsection{Hybrid methods}\label{subsubsec3}

\begin{table}[!t]
\begin{center}
\caption{Summary of hybrid methods in semi-supervised medical multi-organ segmentation}\label{tab4}%
\footnotesize
\begin{tabular}{@{}m{1cm}<{\centering}m{1.3cm}<{\centering}m{1cm}<{\centering}m{4cm}<{\centering}m{3cm}<{\centering}m{3cm}<{\centering}@{}}

\toprule
Reference & Segmentation mode & Site    & Generation \& Utilization of pseudo labels                                                  & Supervision way                                                     & Constraint                                    \\
\hline

\cite{luo_semi-supervised_2021} & Global            & Heart   & Dual parallel networks of Transformer and CNN                                               & Output-level cross pseudo supervised                                & Consistency between two networks              \\
\cite{wu_mutual_2022} & Global            & Heart   & Shared-encoder and multi-decoders network; uncertainty guidance                             & Output-level consistency loss                                       & Mutual consistency of multiple decoders       \\
\cite{liu_semi-supervised_2022} & Global            & Heart   & Shape-agnostic Network \& shape-aware Network; uncertainty guidance                         & Output-level cross pseudo supervised                                & Shape awareness and local context constraints \\
\cite{wang_semi-supervised_2022} & Global            & Heart   & Teacher-student network with single encoder and triple decoders; Tiple-uncertainty guidance & Output-level consistency of Mean Teacher \& feature-level contrastive loss          & Contrastive learning \& consistency constraint                       \\
\cite{chen_mass_2022} & Global            & Abdomen & Modality-specific networks                                                                  & Output-level consistency \& feature-level contrastive loss         & Cross-modality consistency constraint                   \\
\cite{wu_exploring_2022} & Global            & Heart   & Single U-Net segmentation network                                                           & Output-level smoothness loss \& feature-level class-separation loss & Class and smoothness constraint               \\
\cite{lin_calibrating_2022} & Associative       & Knee    & Two parallel segmentation models                                                            & Uncertainty sampling supervision                                    & Distribution-based sampling constraint        \\
\cite{chen_uncertainty_2022} & Global            & Heart   & Uncertainty-aware mean teacher; Uncertainty-based label assignment                          & Output-level consistency                                            & Focal loss for multi-organ   \\
\bottomrule
\end{tabular}
\end{center}
\end{table}

Naturally, the aforementioned methods are not antagonistic but can be combined organically to facilize the performance, while more additional constraints can be included in the semi-supervised learning, which is all regarded as hybrid methods as follow. Table \ref{tab4} lists some hybrid semi-supervised multi-organ segmentation methods.

It's easy to combine the idea of the pseudo labels and consistency regularization for the consistency of different pseudo labels should be ensured under perturbations. Thus, it comes to the idea that the pseudo labels can be exchanged between two networks to realize cross supervision, that is cross pseudo supervised (CPS)~\cite{chen_semi-supervised_2021}, the structure of which is shown in Figure 3 (b). Followed by CPS, Luo et al.~\cite{luo_semi-supervised_2021} combined the hot-spot Transformer~\cite{dosovitskiy_image_2021} by forming a two-branch framework with CNN to ally the issue of the limited perceptual field of CNN by the powerful global information extraction of the Transformer. Wu et al.~\cite{wu_mutual_2022} proposed MC-Net+ developing a mutual consistency to encourage an invariant prediction result of slightly different multiple decoders in easily misclassified regions (e.g., adhesive edges or thin branches), in which the statistical discrepancy of multiple decoders’ outputs is used to estimate the model’s epistemic uncertainty denoting the segmentation difficulty of regions. And the mutual consistency constrain between one decoder's probability output and other decoders' soft pseudo labels. 

Except for consistency constraints, other kinds of constraints also do good to improve the pseudo labels' quality, like the work by Liu et al.~\cite{liu_semi-supervised_2022} using shape awareness and local context constraints. Furthermore, the uncertainty generated by Monte-Carlo dropout (MC Dropout)~\cite{gal_dropout_2016} is included to guide better utilization of the pseudo labels. Besides, Wang et al.~\cite{wang_semi-supervised_2022} introduced contrastive learning on the feature level beyond their proposed task-level consistency to explore a more powerful representation. By adding two auxiliary tasks, i.e., a reconstruction task for capturing semantic information and a signed distance field (SDF)~\cite{xue_shape-aware_2020} regression task, it can impose shape constraint to the segmentation, and these three tasks’ mutual promotion effect is explored under mean teacher architecture and guided by triple-uncertainty via uncertainty weighted integration strategy to learn more reliable knowledge from unlabeled data. Another use of contrastive learning, Chen et al.~\cite{chen_mass_2022} exploited cross-modal consistency to achieve a modality-collaborative semi-supervised segmentation, which leverage the modality-specific knowledge learned from unpaired CT and MRI, and the contrastive similarity is used to regularize the feature extraction. Also constraint on the feature level, Wu et al.~\cite{wu_exploring_2022} proposed SS-Net to realize effective semi-supervised performance through pixel-level smoothness constraints under adversarial perturbations and inter-class separation based on prototype strategy.

Lin et al.~\cite{lin_calibrating_2022}, then, focus on the class imbalance problem of multiple organs, which has a more prominent impact on semi-supervision. They set the category-aware loss function and oversampled cropping region~\cite{roth_hierarchical_2017} based on the number and probability distribution of labels, and use uncertainty sampling supervision to augment the supervision of low-confidence categories and optimize segmentation. Chen et al.~\cite{chen_uncertainty_2022} introduced a dense focal loss~\cite{ma2021loss} function on deep co-training to solve the class imbalance problem and avoid erroneous optimization by an uncertainty-based label assignment policy.

\subsection{Fusing partially annotated datasets - Partially-supervised learning}\label{subsec3}

\begin{figure}
\centering
\includegraphics[width=1.0\textwidth]{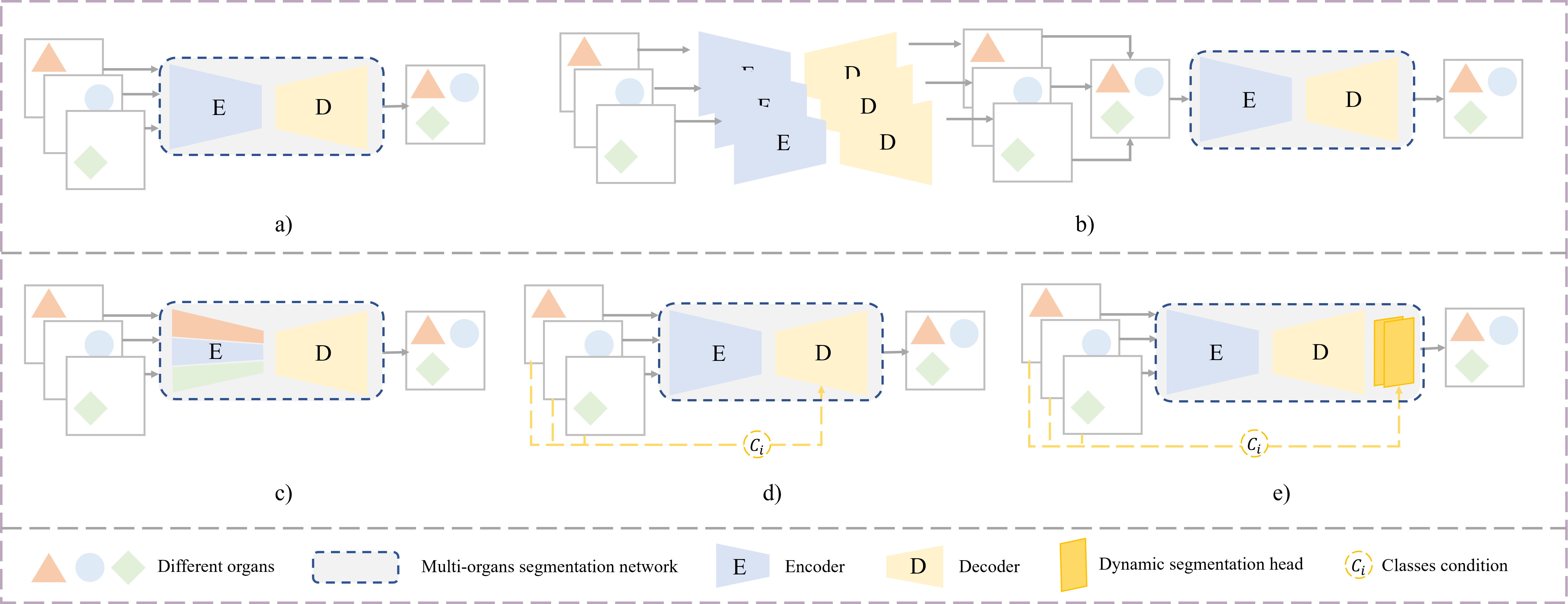}
\caption{Frameworks of partially-supervised learning in multi-organ segmentation. The first row presents two kinds of unified training frameworks, where the same multi-organ segmentation network is trained jointly with different datasets. The second row presents three kinds of separative training, where different data flows through slightly different multi-organ segmentation network which will change with different classes.}\label{<figure4>}
\end{figure}

Due to the enormous labor required for multi-organ annotations, partially annotated multi-organ datasets are more readily available than fully annotated multi-organ datasets. For example, KiTS~\cite{heller_kits19_2019} and LiTS~\cite{bilic_liver_2019} datasets only provided labels for a single organ and tumors in the abdomen, while all other organs are marked as background. Learning from multiple datasets containing single or several organs' annotations together can integrate multiple partial annotations to provide full annotation for desired multiple organs, and such a learning approach is called partially-supervised learning. Partially supervised learning can also be regarded as a kind of the transfer learning or semi-supervised learning, where learning from multiple partially-labeled datasets involves knowledge transfer learning among different datasets, while unlabeled anatomical structures in partially-labeled data  may serve as unlabeled data for semi-supervised learning. Partially supervised learning aims to generate a framework universal to multi-organ segmentation from partially-labeled data. The organs of interest may be marked as background in partially-labeled datasets, which may mislead and confuse the multi-organ segmentation. For this reason, how to make better utilization of the partially labeled information without confusing the overall multi-organ segmentation is a key issue for partial supervision. Based on the way to incorporate the partially-labeled datasets in training, we spilt the partial learning methods into unified training and separate training, which are distinguished by the frameworks in Figure \ref{<figure4>}. The upper row shows the unified training in which the data of different partially-annotated datasets flow through the same multi-organ network while in the separative training showed in the second-row the data flow through the multi-organ network differentially. In the former case, different datasets jointly update the whole network parameters, while in the latter case the optimization of the parameters by different datasets is local or separated. Table \ref{tab5} lists the partially-supervised learning methods in multi-organ segmentation.

\subsubsection{Unified training}\label{subsubsec1}

To obtain a multi-organ segmentation network, the partially-labeled datasets can be fed into a unified model in which the multiple partially-labeled datasets share the same training architecture. In such unified training, the training with each dataset will benefit from  same designed partial supervised loss. Shi et al.~\cite{shi_marginal_2021}  exploited the union of partially labeled datasets through two carefully-designed loss functions, where a marginal loss mines the true background from partially-labeled datasets through the boundary probabilities, and an exclusion loss was designed to improve inter-organ discrimination considering the fact that there is no overlap between organs. Both of these loss functions showed apparent improvement to the segmentation performance without introducing any additional computation. With the aim of further utilizing the contextual information, Mao et al.\cite{hongdong_multi-scale_2022} extended these loss functions and imported attention mechanism to collect multi-scale information for fully automatic multi-organ segmentation, which better strikes a trade-off between segmentation accuracy and inference time. Fang et al.~\cite{fang_multi-organ_2020} proposed a unified U-shape pyramid structure network that aggregates pyramid input and features where an adaptive weighting layer is designed to integrate the deeply-supervised outputs in an automatic way. This method allows a single model adapt to changes in proportion of known labels via a target adaptive loss.

Additionally, using pseudo labels is also a feasible manner for unified training. Zhou et al.~\cite{zhou_prior-aware_2019} proposed PaNN using very little fully-labeled data for training to obtain a prior-aware network, and then optimized it by using the partially-labeled data by the supervised loss computed on labeled organs and the pseudo label-based loss on unlabeled organs. PaNN regularizes the organ sizes by explicitly incorporating anatomical empirical distributions as the penalty of soft constraint with statistical domain-specific knowledge. Huang et al.~\cite{huang2020multi} deployed two coupled networks for few-organ datasets to teach each other and proposed a co-train weight-averaged model to obtain a unified multi-organ segmentation network. The multiple organs' pseudo labels generated by two coupled networks can be fused to form a fully multi-organ pseudo labels, which can further refine the unified segmentation network. In addition, this work proposes a novel region mask to selectively impose the consistent constraint on the unlabeled organ regions during collaborative teaching, thus delivering further performance improvements. 

In addition, Zhou et al.~\cite{zhou_uncertainty-aware_2021} considered that the partially-labeled dataset may not be publicly available in real scenarios, so they unleash the partially annotated and sequentially constructed datasets through incremental learning (IL) without access to previous annotations in each IL stage. Pre-trained on a public dataset of K organs, the model is trained with another single-organ dataset to obtain a segmentation of K+1 organs via IL, and apply a teacher-student framework to transfer knowledge extracted by background remodeling. However, the unified training may be affected by the distribution gap of different datasets resulting a mutual interference.

\begin{table}[!t]
\begin{center}
\begin{minipage}{\textwidth}
\caption{Summary of partially supervised medical multi-organ segmentation methods}\label{tab5}%
\footnotesize
\begin{tabular}{@{}m{1cm}<{\centering}m{1.5cm}<{\centering}m{1.4cm}<{\centering}m{4.5cm}<{\centering}m{4cm}<{\centering}m{1.5cm}<{\centering}@{}}

\toprule
Reference & Segmentation mode & Multi-organ dataset & Partially labeled datasets                                 & Integration of partial labelling                                  & Framework \\
\midrule
\cite{zhou_prior-aware_2019}  & Global            & BTCV~\cite{landman_miccai_2015}      & MSD Spleen~\cite{simpson_large_2019}, NIH~\cite{roth_data_2016}, LiTS~\cite{bilic_liver_2019}        & A priori soft constraints on size                                 & a         \\
\cite{shi_marginal_2021}  & Global            & BTCV~\cite{landman_miccai_2015}      & MSD Spleen~\cite{simpson_large_2019}, KiTS~\cite{heller_kits19_2019}                                 & Marginal and excluded losses                                      & a         \\
\cite{hongdong_multi-scale_2022} & Multi-scale       & PDDCA\cite{raudaschl_evaluation_2017}      & TCIA-CT-scan~\cite{clark_cancer_2013,vallieres_radiomics_2017}                               & Attentional mechanisms integrate multi-scale information          & a         \\
\cite{fang_multi-organ_2020} & Multi-scale       & BTCV~\cite{landman_miccai_2015}       & LiTS~\cite{bilic_liver_2019}, KiTS~\cite{heller_kits19_2019} , MSD Spleen~\cite{simpson_large_2019}                & Adaptive fusion \& adaptive loss                                  & a         \\
\cite{huang2020multi} & Nested            & $\times$                   & LiTS~\cite{bilic_liver_2019}, KiTS~\cite{heller_kits19_2019} , NIH~\cite{roth_data_2016}, MOBA~\cite{gibson_automatic_2018} & Pseudo-label generation \& Co-training and weight- averaged model & b         \\
\cite{zhou_uncertainty-aware_2021} & Global            & MOBA~\cite{gibson_automatic_2018}        & MSD Spleen~\cite{simpson_large_2019}, LiTS~\cite{bilic_liver_2019}, NIH~\cite{roth_data_2016}             & Knowledge distillation                                            & a         \\
\cite{xu_federated_2022} & Associative       & $\times$                   & LiTS~\cite{bilic_liver_2019}, KiTS~\cite{heller_kits19_2019} , MSD Spleen~\cite{simpson_large_2019}, BTCV~\cite{landman_miccai_2015}     & Organ-individual encoders and shared Decoder                      & c         \\
\cite{dmitriev_learning_2019} & Global            & $\times$                   & Sliver07~\cite{heimann_comparison_2009}, NIH~\cite{roth_data_2016}, liver \& spleen*            & Introduction of categories condition                              & d         \\
\cite{zhang_automatic_2021} & Associative       & $\times$                   & LITSC*, MSD LT*                                             & Semi-supervision and conditional strategy                         & d         \\
\cite{zhang_dodnet_2021} & Associative       & $\times$                   & MOTS~\cite{zhang_dodnet_2021}, BTCV~\cite{landman_miccai_2015}                                & Task-encoding controller                                          & e       \\
\bottomrule
\end{tabular}
\end{minipage}
\end{center}
\end{table}

\subsubsection{Separative training}\label{subsubsec2}

However, influenced by the previous background confusion problem, training partially-labeled data indiscriminately with a simple single model is not an optimal way to fuse different datasets.

Thus, Xu et al.~\cite{xu_federated_2022} explored a multi-encoding U-Net to extract organ-specific features through different several encoding sub-networks, and implement federated multi-organ segmentation by an auxiliary generic decoder. This approach explores the informative and distinctive information for different organs by multiple encoders but balances the resource with a shared decoder. However, such a design may rise the redundancy and memory usage of the network. 

Taking into account the need to reduce redundancy, the use of conditional strategies in the union network allows the introduction of category conditions as auxiliary information for the segmentation of different organs to achieve differentiated segmentation. Inspired by conditional generative networks, Dmitriev et al.~\cite{dmitriev_learning_2019} utilized a conditional approach to gain a joint multi-organ framework trained from partially-labeled datasets. It is simple yet efficient and shows convincing segmentation performance on multi-organ images. Likewise, Zhang et al.~\cite{zhang_automatic_2021} regard two datasets with non-overlapping labeling pairs as mutually unlabeled data and use the classic semi-supervised teacher-student framework to learn from a collection of partially-labeled datasets. The backbone is built on the nnU-Net combined a conditioning strategy to include prior class information into the decoder to better utilize relative unlabeled data, i.e., partially-labeled data.

Going further, Zhang et al.~\cite{zhang_dodnet_2021} proposed a dynamic on-demand network (DoDNet) by the generated controller, where the majority of the network is a shared encoder-decoder architecture with dynamic segmentation head changed with the class of organ. Different from fixing the kernels after training from the previous works, DoDNet generates the kernels in dynamic head adaptively by the task encoding module, enabling segmentation of multiple organs and tumors in a more efficient and flexible manner. However, the method cannot achieve simultaneous segmentation of multiple organs and can only perform one-by-one inference. Currently, there is still room to explore a flexible, efficient, and low-resource segmentation framework with the integration of partially labeled data in partially supervised tasks.

\section{Discussion and future prospects}\label{sec4}

Current deep learning approaches have already served an important role in multi-organ segmentation, which is significant in helping to reduce annotation consumption and alleviate inter-observer annotation discrepancies. In this paper, we systematically demonstrate the multi-organ segmentation framework and explore the latest works with the data scarcity problem as the entry point. The present annotation-efficient algorithms have shown effective clinical practicability in reducing annotation cost in the field of multi-organ segmentation, however, how to further develop more high-accuracy, low-resource and annotation-efficient segmentation algorithms for multi-organ segmentation still need to be explored.

\subsection{Transfer learning}

Transfer learning aims to make full use of external annotated data to enhance the performance of multi-organ segmentation, and the key lies in solving domain shift between datasets. Transfer learning currently focuses on the migration of homogeneous data across data sources and adaptation between different modalities, while the cross-modality transfer towards functional imaging with greater differences is still lacking for there is no guarantee to capture sufficient features for cross-modality generation with enormous differences in tissue presentations~\cite{jiang_unpaired_2022}. Nevertheless, the functional imaging presents lesion region better making the transfer learning with functional images essential for precise segmentation of tumors along with OARs and valuable for further exploration. In addition, embracing artificial interaction in transfer learning can help guarantee the accuracy of target domain segmentation, and using transfer learning to provide a pre-segmentation reference for experts can speed up the efficiency of multi-organ labeling~\cite{ma_rapid_2022}. In the future, scientific research results should concern more about real clinical needs, and more effort should be made to translate the research results into a commercially available online segmentation radiotherapy platform with higher real-time and generalization requirements via transfer learning~\cite{shi_deep_2022,LIANG201834}. Also, the active learning-based approach can be included in transfer learning to select representative samples for annotation thus boosting the adaptation quality and speed~\cite{zheng2020annotation}, which can help construct attempts to build a semi-automatic segmentation platform of high industrial value and clinical feasibility.

\subsection{Semi-supervised learning}

Semi-supervised learning aims to use unlabeled data to help improve the performance of target multi-organ segmentation. Among the common approaches of semi-supervised methods, the pseudo label-based approaches are more general, but require longer training time in multi-stage optimizations and are more sensitive to the quality of pseudo labels; the consistency regularization-based approaches enable flexible and convenient construction of regularization constraints, but the tight coupling constructed through consistency may cause bottlenecks in accumulated bias correction. And the above issues have been explored by some corresponding solutions~\cite{hu_semi-supervised_2022,ke2019dual}, however, little work has been done to focus on class imbalance in multi-organ segmentation. So how to effectively improve the quality of pseudo labels of tiny organs or, alternatively, to take into account both global and local segmentation from a consistency perspective are directions that can be investigated in the future. In addition, the few-shot learning~\cite{wang_generalizing_2020} approach, similar to semi-supervised learning has also been applied in multi-organ segmentation tasks~\cite{hansen2022anomaly}, reflecting a more critical demand for more annotation efficiency. Moreover, similar to the enhancement of representation in unsupervised learning, the use of self-supervised paradigm~\cite{chen2020simple,he2020momentum}, e.g., contrastive learning to learn the differences between different organs, etc.~\cite{liu_context-aware_2022,wang2022separated} can help networks better extract valid information from unlabeled data and is likewise worth exploring in semi-supervised learning.

\subsection{Partially-supervised learning}

Most methods in partially supervised learning tend to focus on how to construct an effective multi-organ segmentation network. To obtain such multi-organ segmentation network, either unified or separative training, it is crucial to exclude interference between different partially labeled datasets, but it is more difficult to effectively establish contextual information links between different organs when full annotations are scarce~\cite{hongdong_multi-scale_2022}. Partially supervised learning can also be regarded as a kind of semi-supervised learning or transfer learning, where unlabeled anatomical structures in partially-labeled data can be employed may serve as unlabeled data for semi-supervised learning, while learning from multiple partially-labeled datasets involves knowledge transfer learning among different datasets. And it’s inevitable to face the same challenges of these two learning for the partially-supervised learning. Direct use of global pseudo labels can help achieve a contextual constraint, but it is difficult to balance the delineation performance of organs with various sizes~\cite{fang_multi-organ_2020}. How to introduce anatomical information constraints in a joint unified multi-organ segmentation framework is still worth exploring. Considering the discrepancies between partially labeled datasets, the existing methods give the assumption of basic consistent data distribution and standardize the data through complex preprocessing. As different datasets continue to be introduced and used, how to adaptively learn knowledge from new samples and preserve most of the previously learned knowledge is also a subsequent trend in fusing partially labeled data~\cite{zhou_uncertainty-aware_2021}.

\section{Conclusion}\label{sec5}

In this paper, we provide a systematic summary of multi-organ segmentation from fully supervised to annotation-efficient learning paradigms, and unfold the current state of works in transfer learning, semi-supervised learning, and partially supervised learning. Further, we provide a general review and outlook on the problems and shortcomings of these approaches. The current annotation-efficient multi-organ segmentation approaches are under a period of rapid development, with smaller annotation requirements, wider types of applications, and increasing prediction accuracy. Future exploration of more efficient automatic multi-organ segmentation is also expected to reduce the medical burden in areas with insufficient medical resources~\cite{men_automatic_2017} and benefit cancer patients, which has far-reaching implications.






\begin{thebibliography}{100}
\bibitem{tarvainen2017mean}
  A.~Tarvainen, H.~Valpola, Mean teachers are better role models: Weight-averaged
    consistency targets improve semi-supervised deep learning results, Advances
    in neural information processing systems 30.
\end{thebibliography}







\end{document}